\documentclass[pdftex,twocolumn,epjc3]{svjour3}          
\RequirePackage[T1]{fontenc}
\smartqed  
\RequirePackage{graphicx}
\RequirePackage{mathptmx}      
\RequirePackage{flushend}
\RequirePackage[numbers,sort&compress]{natbib}
\RequirePackage[colorlinks,citecolor=blue,urlcolor=blue,linkcolor=blue]{hyperref}

\usepackage{amssymb,dcolumn}
\usepackage{amsmath,lipsum}
\usepackage{xcolor}
\usepackage{multirow}
\usepackage{extarrows}
\usepackage{float}
\usepackage{graphicx,subfigure}
\usepackage{bm}
\usepackage{enumerate}
\usepackage{mathrsfs}
\usepackage{microtype}
\usepackage[english]{babel}
\usepackage{multicol}
\usepackage{cuted}
\usepackage{changepage}

\journalname{Eur. Phys. J. C}

\begin{document}

\title{Shadow and Optical Imaging in Einstein-Maxwell-Dilaton Black Hole}

\author{Junlin Qin\thanksref{e1,addr1}
        \and
        Hong-Er Gong\thanksref{e2,addr1}
        \and
        Yusen Wang\thanksref{addr1}
        \and
        Zhan-Feng Mai\thanksref{e3,addr1}
        \and
        Bofeng Wu\thanksref{e4,addr1}
        \and
        Sen Guo\thanksref{addr2}
        \and
        Enwei Liang\thanksref{e5,addr1}
}
\thankstext{e1}{e-mail: junlin.qin@st.gxu.edu.cn}
\thankstext{e2}{e-mail: gonghonger@yeah.net}
\thankstext{e3}{e-mail: zf1102@gxu.edu.cn (corresponding author)}
\thankstext{e4}{e-mail: wubofeng@gxu.edu.cn (corresponding author)}
\thankstext{e5}{e-mail: lew@gxu.edu.cn}

\institute{Guangxi Key Laboratory for Relativistic Astrophysics, School of Physical Science and Technology, Guangxi University, Nanning 530004, China\label{addr1}
          \and
          College of Physics and Electronic Engineering, Chongqing Normal University,
          Chongqing 401331, China\label{addr2}
}

\date{Received: date / Accepted: date}
\maketitle

\begin{abstract}
   This paper investigates photon motion in black hole of Einstein-Maxwell-dilaton theory, exploring black hole shadows and observational characteristics under various accretion models. We first give the relation of the event horizon, photon sphere, and critical impact parameter in terms of the magnetic charge $q$. We then use the Event Horizon Telescope data to constrain $q$. For the two spherical accretion models, the infalling scenario yields a darker shadow due to the Doppler effect. However, the shadow radius remains unchanged for different models. In the case of an optically thin, geometrically thin disk accretion model, the observed brightness is predominantly determined by direct emission. The lensing ring provides a secondary contribution to the intensity, whereas the photon ring's emission is negligible. The widths of the lensing and photon rings exhibit a positive correlation with the magnetic charge $q$. Additionally, within the disk model framework, the black hole shadow radius is found to depend on the specific emission model. 
\end{abstract}
\newpage

\section{Introduction}
Black holes are extremely compact celestial objects predicted by General Relativity. Since the first black hole solution was derived from the field equations, scientists have been searching for observational evidence of these mysterious objects in the universe \cite{r1}. The successful detection of gravitational waves from binary black hole mergers by the Laser Interferometer Gravitational-Wave Observatory (LIGO), together with the discovery of massive star-black hole binary systems via radial velocity measurements, have robustly confirmed the existence of black holes \cite{r2,r3,r4,r5}. The intense gravitational field of a black hole causes photon geodesics to deviate from straight-line trajectories, resulting in light deflection and the formation of black hole shadows—a phenomenon known as gravitational lensing \cite{r6,r7,r8,r9,r10}. Subsequently, the Event Horizon Telescope (EHT) Collaboration captured the first image of the supermassive black hole at the center of the M87 galaxy \cite{r11,r12,r13,r14,r15,r16,r17,r18,r19,r20,r21,r22,r23,r24,r25,r26,r27}, providing compelling evidence for their existence. This image reveals a bright ring surrounding a central dark region. The dark region is identified as the black hole shadow, whereas the bright region corresponds to the photon ring \cite{r28}.

It is widely accepted that black holes constitute the final stage of gravitational collapse for massive stars. Given that these stars and their surrounding material possess angular momentum, the resulting black holes inherit a significant portion of it from their progenitors. As a consequence, most of cosmic black holes are rotating. Bardeen was the first to analytically caclulate the shadow of a Kerr black hole, observing that frame-dragging in rotating spacetimes distorts the shadow from the standard circular form produced by the lensing ring into a "D" shape \cite{r29}. Additionally, the rotational energy of the rotating black hole can be extracted via the Penrose process, the superradiance effect \cite{r30} or the Blandford–Znajek mechanism \cite{r31}, ultimately leading to their degeneration into non-rotating black holes. Thus, investigating the optical characteristics and conducting image simulations for static, spherically symmetric black holes remains highly relevant.

In static, spherically symmetric spacetimes, black hole shadows exhibit a standard circular geometry due to the gravitational lensing effect \cite{r32}. Bardeen established that the shadow radius of a Schwarzschild black hole, $r_s=3\sqrt{3}M$ (where $M$ is the black hole mass) \cite{r29}. Narayan et al. simulated spherical accretion onto Schwarzschild black holes, finding that the shadow boundary remains invariant regardless of the initial position of accreting matter \cite{r33}. Q.-Y. Gan et al. investigated spherical accretion for hairy black holes, revealing the presence of two photon rings and identifying the smaller one as the shadow boundary \cite{r34}. Wu et al. employed a semi-analytical method to simulate a more comprehensive optical image of Schwarzschild black holes \cite{r35}. Furthermore, Gong et al. compared the shadows and photon rings of two minimally deformed Schwarzschild black holes \cite{r35.5}.

In optical simulations of black hole, spherical accretion offers an idealized depiction of the matter accreting near a black hole. In fact, due to the intense gravitational pull, infalling matter forms a disk-like structure around the black hole, known as an accretion disk. For a distant observer, radiation emitted from this accretion disk travels across space, rendering the black hole's silhouette visible. Building upon Luminet's work on rotating, geometrically thick accretion disks \cite{r36}, Gralla et al. pioneered the simulation of shadow imaging for a Schwarzschild black hole surrounded by an optically thin, geometrically thin accretion disk, identifying the presence of photon rings and lensing rings within the shadow \cite{r37}. Owing to the limited angular resolution of the EHT, these rings cannot be directly resolved under specific thin-disk accretion conditions. Consequently, researchers have subsequently performed extensive studies on shadow imaging of black holes with thin-disk accretion \cite{r38,r39,r40,r41}.

String theory, a leading candidate for a quantum theory of gravity, naturally couples the dilaton field to Einstein-Maxwell fields in its low-energy limit and via dimensional reduction from five-dimensional Kaluza-Klein theory \cite{r42}. EMD gravity extends the standard Einstein-Hilbert action by  Maxwell field and a scalar field (the dilaton field), thereby creating avenues for exploring quantum gravity \cite{r43}. Research on black holes in EMD theory has been extensive. For instance, studies in \cite{r44,r45} investigate charged, static, spherical black hole within this framework, while \cite{r46} generalizes these static solutions to rotating black holes. Notably, Wu et al. analyze the properties of EMD black holes as a specific case of EMS black holes with $\beta=0$ \cite{r47}. Furthermore, \cite{r48} compares EMD black holes with Reissner-Nordstr\"{o}m (RN) black holes. Chen et al. investigates the shadows of EMD black holes \cite{r48.5}.

This paper is organized as follows: In Sec.~\ref{sub2}, we derive the metric of EMD gravity action, constrain its parameters using EHT observations and present the light trajectories for different black hole. In Sec.~\ref{sub3}, we discuss the optical characteristics of the EMD black holes under both static and infalling spherical accretion model. In Sec.~\ref{sub4}, we investigate the observational implications of various emission models on EMD black holes, assuming an optically thin and geometrically thin accretion disk. We summarize
and discuss our results in Sec.~\ref{sub5}. This work adopts geometrized units, where \(c = G = 1\).

\section{Spacetime of EMD Black Hole\label{sub2}}
In this paper, we consider the static spherical ansatz \begin{align}
     \mathrm{d}s^2 = -A(r) \mathrm{d}t^2 + B(r) \mathrm{d}r^2+C(r) \mathrm{d}\Omega^2,
     \label{ds^2}
\end{align}
where $\mathrm{d}\Omega^2=\mathrm{d}\theta^2+\sin^2\theta\mathrm{d}\varphi^2$. For the EMD BH, the Lagrangian takes\cite{r48}
\begin{align}
     \mathscr{L}=e(-R + 2(\nabla\phi)^2 + e^{-2\phi}F^2),
\end{align}
where $e$ is the determinant of the metric. In addition, $R$ is the Ricci scalar and $\phi$ is the dilaton field. $F^2=\frac{1}{4}F^{\mu\nu}F_{\mu\nu}$ is the field strength of electromagnetic field. $F_{\mu\nu}=\partial_\mu A_\nu - \partial_\nu A_\mu$ is the electromagnetic field tensor in which $A_\mu$ is the electromagnetic four-potential. Consider the variation of the EMD Lagrangian with respect to $A_\mu$, $\phi$, and $g^{\mu\nu}$, respectively, yields \cite{r48}
\begin{equation}
    \begin{aligned}
        & \nabla_\mu (e^{-2\phi}F^{\mu\nu}) = 0, \\
        & \nabla^2\phi+\frac{1}{2}e^{-2\phi}F^2=0, \\
        & R_{\mu\nu} = 2\nabla_\mu\phi \nabla_\nu\phi + 2e^{-2\phi}F_{\mu\rho}F_\nu^{\ \rho}-\frac{1}{2}g_{\mu\nu}e^{-2\phi}F^2.
        \label{L}
    \end{aligned}
\end{equation}
Substituting Eq.~\eqref{ds^2} into the field equations Eq.~\eqref{L} allows for obtaining the solution, while imposing the requirements that the solution is asymptotically flat and possesses a regular horizon
\begin{equation}
        \begin{aligned}
            A(r) &= \frac{1}{B(r)}= 1-\frac{2M}{r},\\
            C(r) &= r(r-\frac{q^2}{M}),
        \end{aligned}
\end{equation}
where $q$ can be viewed as the magnetic charge, which is different from the electric charge. In this paper, it is characterized as a form of ``hair'' for the EMD black hole \cite{r49, r50}. Since $C(r) \neq r^2$ differs from conventional Schwarzschild spacetime, the coordinate $r$ does not correspond to the effective radial distance. Consequently, we apply a coordinate transformation satisfying $C(r) = r^2$, yielding \cite{r51}
    \begin{equation}
        \begin{aligned}
            A(r) &= 1-\frac{4M}{\chi+\sqrt{\chi^2 + 4r^2}}, \\
            B(r) &= \frac{4r^2(\chi+\sqrt{\chi^2+4r^2})}{(\chi^2+4r^2)(\chi-4M+\sqrt{\chi^2+4r^2})},\\
            C(r) &= r^2,
            \label{C(r)}
        \end{aligned}
    \end{equation}
where $\chi=q^2/M$. If we consider the case of $q=0$,  one will find  $A(r) = \dfrac{1}{B(r)} = 1-\dfrac{2M}{r}$ and $C(r) = r^2$, which reduces to the Schwarzschild BH. Furthermore, when $\chi$ is small, the metric function $A(r)$ of the EMD black hole is very close to that of the RN black hole, leading to similar appearances in the subsequent simulations. The difference is that for the RN black hole, $A(r)=\dfrac{1}{B(r)}$, whereas for the EMD black hole, $A(r)\neq\dfrac{1}{B(r)}$.

For the SC BH, RN BH, and EMD BH, their respective event horizons can be determined by solving the equation $A(r) = 0$,
\begin{itemize}
    \item SC BH\begin{align}
        r_h = 2M,
    \end{align}
    \item RN BH\begin{align}
        r_h = M\pm\sqrt{M^2-Q^2},
    \end{align}
    \item EMD BH\begin{align}
        r_h = \sqrt{4M^2-2q^2}.
        \label{rhEmdg}
    \end{align}
\end{itemize}
The magnetic charge $q$ is negatively correlated with the event horizon $r_h$ of the EMD black hole. Therefore, it is important to impose restrictions on the value of $q$. Otherwise, it would violate the cosmic censorship conjecture, leading to the appearance of a naked singularity (the constraints on the range of $q$ from EHT are provided later).

According to \cite{r52}, we have
\begin{align}
    C'(r_{p})A(r_{p})-C(r_{p})A'(r_{p})=0 ,\qquad
        b_{p} = \sqrt{\frac{C(r_{p})}{A(r_{p})}}.
\end{align}
compared to the SC BH and RN BH, the photon sphere radius $r_{p}$ and the critical impact parameters $b_{p}$ for the three black holes are as follows, (hereafter, we set $M=1$)
\begin{itemize}
    \item SC BH\begin{align}
        r_{p} &= 3,\\
        b_{p} &= 3\sqrt{3}.
    \end{align}
    \item RN BH\begin{align}
        r_p &= \frac{3}{2}\pm\frac{1}{2}\sqrt{9-8Q^2},\\
        b_p &= \sqrt{\frac{(3\pm\sqrt{9-8Q^2})^4}{8(3-2Q^2\pm\sqrt{9-8Q^2})}}.
    \end{align}
    \item EMD BH
    \begin{equation}
        r_{p} = \frac{\sqrt{36-16q^2-q^4+6\sqrt{a}-q^2\sqrt{a}}}{2\sqrt{2}},
    \end{equation}
\begin{equation}
    \begin{aligned}
        b_{p} &= \frac{1}{2}\{[(-2+q^2)\left(q^6+q^4\left(2+\sqrt{a}\right)\right.\\
        &\left.\left.-108\left(6+\sqrt{a}\right)+12q^2\left(21+\sqrt{a}\right)\right)\right]\\
        &/ [36-q^4+6\sqrt{a}-q^2\left(8+\sqrt{a}\right)\\
        &\left.\left. -4\sqrt{2}\sqrt{q^4+6\left(6+\sqrt{a}\right)-q^2\left(16+\sqrt{a}\right)}\right]\right\}^\frac{1}{2}.
        \label{bpEMdg}
    \end{aligned}
\end{equation}
\end{itemize}
where $a = 36-20q^2+q^4$. Table \ref{rhrpbp} presents the values of $r_h$, $r_p$, and $b_p$ for the EMD BH under different parameters.
\begin{table}[ht]
    \centering
    \setlength{\tabcolsep}{5pt}     
         \caption{Under the condition of mass $M=1$, the values of the event horizon, photon sphere radius, and critical impact parameter for different values of the magnetic charge $q$.}
        \vspace{0cm}                  
        \begin{tabular}{c c c c c c c}  
            \hline
            $q$ & 0 & 0.2 & 0.4 & 0.6 & 0.8 & 1\\
            \hline
            $r_h$   & 2     & 1.97990 & 1.91833 & 1.81108 & 1.64924 & 1.41421 \\
            $r_p$   & 3     & 2.97321 & 2.89123 & 2.74877 & 2.53462 & 2.22530 \\
            $b_p$   & 5.19615 & 5.16136  & 5.05503  & 4.87069 & 4.59486 & 4.19960 \\
            \hline
        \end{tabular}
        \label{rhrpbp}    
\end{table}

Let the photon's four-momentum be $K^\mu=\mathrm{d}x^\mu/\mathrm{d}\lambda$, where $\lambda$ represents the affine parameter along the photon's worldline. From Eqs. \eqref{ds^2} to \eqref{C(r)}, it is apparent that the EMD metric exhibits independence from both $t$ and $\varphi$. Consequently, the spacetime admits two Killing vector fields: $\xi_1^\mu = (\dfrac{\partial}{\partial t})^\mu$ and $\xi_2^\mu = (\dfrac{\partial}{\partial\varphi})^\mu$, which correspond to two conserved quantities along the photon's trajectory.
\begin{align}
    E := -g_{\mu\nu}K^\mu\xi_1^\nu = A(r)\frac{\mathrm{d}t}{\mathrm{d}\lambda},\quad
    L := g_{\mu\nu}K^\mu\xi_2^\nu = C(r)\frac{\mathrm{d}\varphi}{\mathrm{d}\lambda}.
\end{align}
Without loss of generality, the subsequent discussion assumes $\theta = \dfrac{\pi}{2}$, i.e., motion confined to the equatorial plane. Considering the photon's four-momentum satisfies $g_{\mu\nu}K^\mu K^\nu = 0$, the components of the photon's four-momentum are thus obtained
\begin{align}
    K^\mu = \left(\frac{E}{A(r)},\pm E\sqrt{\frac{1}{A(r)B(r)}-\frac{b^2}{B(r)C(r)}},0,\frac{L}{C(r)}\right).
    \label{K^mu}
\end{align}
We then defined the effective potential function as $V_{\mathrm{eff}} := \sqrt{\dfrac{A(r)}{C(r)}}$ and the impact parameter $b := \dfrac{L}{E}$.
\begin{figure*}[h]
    \centering
    \begin{subfigure}
        \centering
        \includegraphics[width = 0.4\textwidth]{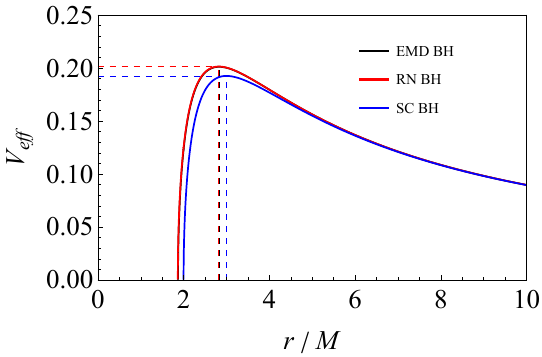}        
    \end{subfigure}
    \hspace{30pt}
    \begin{subfigure}
        \centering
        \includegraphics[width = 0.4\textwidth]{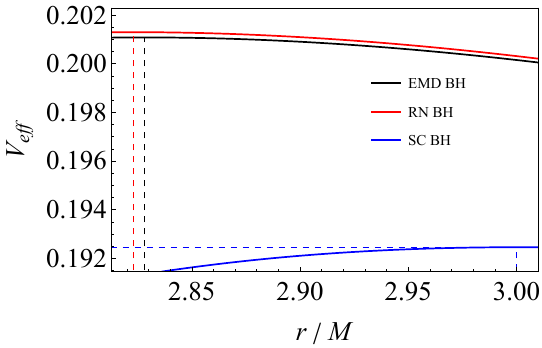}        
    \end{subfigure}
    \caption{Left panel: Effective potential for the SC BH, RN BH, and EMD BH. Right panel: Magnified view of the effective potential near the extremum.}
    \label{Veff}
\end{figure*}

As can be seen from the left panel of Fig. \ref{Veff}, the effective potential originates at the event horizon $r_h$ and reaches an extremum at the photon sphere $r_p$. It is indicated that the magnetic charge $q$ in the EMD BH is negatively correlated with both $r_h$ and $r_p$ in the Table \ref{rhrpbp}, leading to a decrease in the event horizon $r_h$ and the photon sphere radius $r_p$. A smaller photon ring radius indicates a greater curvature of spacetime in the vicinity of the black hole, implying a stronger gravitational field. Since $A(r)$ of the RN BH and the EMD BH are very similar, this results in their effective potentials being very close, as shown in Fig. \ref{Veff}, with only slight differences near the extremum. It is noteworthy that in static spherical spacetimes, the existence of a photon sphere is crucial for the formation of shadow images \cite{r53}. Therefore, using the EHT data for Sgr A*, we can constrain the magnitude of the magnetic charge $q$ in the EMD BH.
\begin{figure*}[h]
    \begin{minipage}{0.48\textwidth}
        \centering
        \includegraphics[width = 0.9\textwidth]{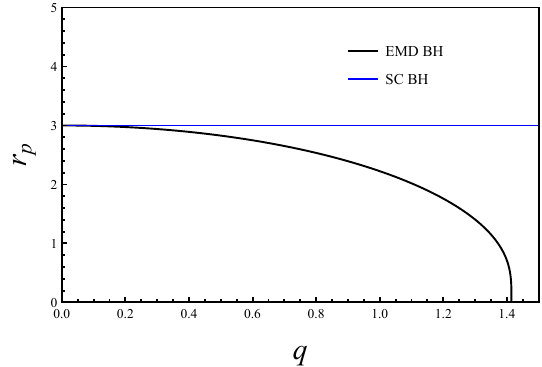}
    \end{minipage}
    \begin{minipage}{0.48\textwidth}
        \centering
        \includegraphics[width = 0.9\textwidth]{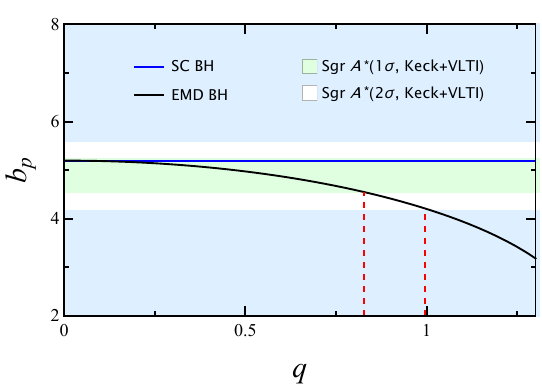}
    \end{minipage}
    \caption{Left panel: The relationship between the photon sphere radius $r_p$ of the EMD black hole and the hair parameter $q$. Right panel: The relationship between the critical impact parameter $b_p$ and the parameter $q$. The right panel utilizes the mass-to-distance ratio prior from EHT observations of $Sgr A^*$, combining data from $Keck$ and $VLTI$ after averaging. The green region represents the 1$\sigma$ confidence interval, the white region represents the 2$\sigma$ interval, and the blue region is excluded as it exceeds 2$\sigma$.}
    \label{EHT}
\end{figure*}

In the left panel of the Fig.\ref{EHT}, we illustrate that $r_p$ exhibits a negative correlation with $q$. As $q$ increases, the photon ring radius decreases rapidly and vanishes at $q=\sqrt{2}$, where the EMD BH reduces to an extremal black hole or a naked singularity. As in the right panel, we utilize observational data from the EHT concerning Sgr A* to constrain the parameters of the EMD BH \cite{r25,r52}. The $1 \sigma$ constraint range is: 4.55 $\lesssim$ $b_p$ $\lesssim$ 5.22, and the $2\sigma$ constraint range is: 4.21 $\lesssim$ $b_p$ $\lesssim$ 5.56. Consequently, the $1 \sigma$ constraint for the EMD BH is: $q$ $\lesssim$ 0.826, and the $2\sigma$ constraint is: $q$ $\lesssim$ 0.995. Therefore, in subsequent sections, the EMD BH with $q=0.5$ is selected for comparison with the SC BH and RN BH (For better comparison with the EMD BH, the charge $Q$ in the RN BH is also set to 0.5.).

\begin{figure*}[h]
    \begin{minipage}{0.32\textwidth}
        \centering
        \includegraphics[width = 0.9\textwidth]{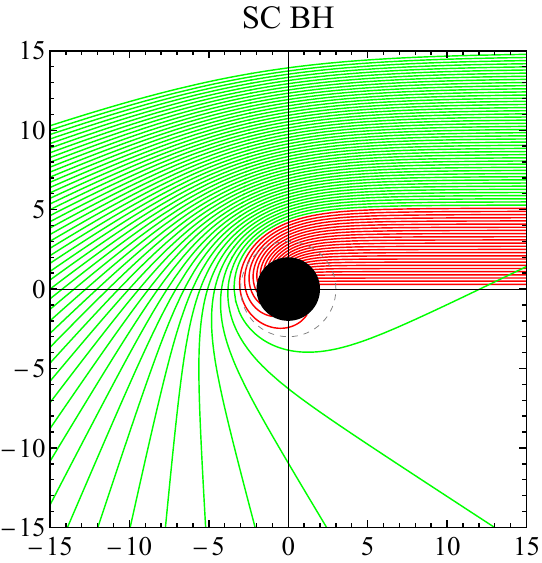}
    \end{minipage}
    \begin{minipage}{0.32\textwidth}
        \centering
        \includegraphics[width = 0.9\textwidth]{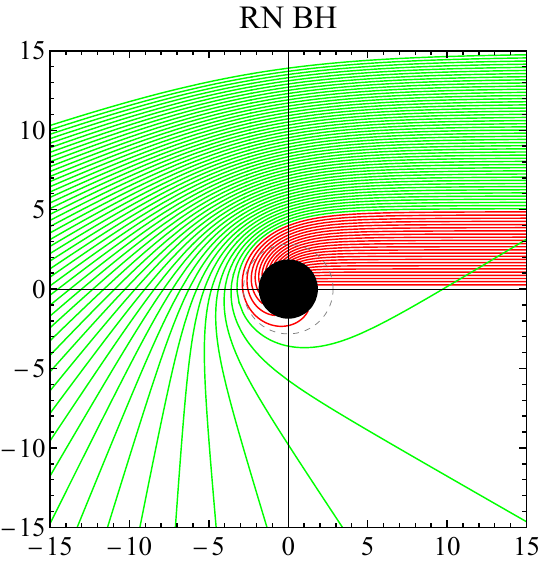}
    \end{minipage}
    \begin{minipage}{0.32\textwidth}
        \centering
        \includegraphics[width = 0.9\textwidth]{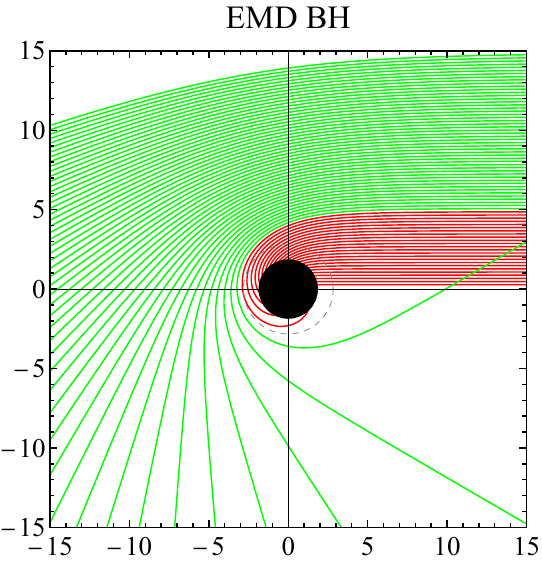}
    \end{minipage}
    \caption{Light ray trajectories in the equatorial plane. From left to right are the SC BH, RN BH, and EMD BH. The black solid circle represents the event horizon, the gray dashed circle represents the photon ring; the red curves correspond to light rays with an impact parameter $b < b_p$; the green curves correspond to light rays with an impact parameter $b > b_p$.}
    \label{light ray1}
\end{figure*}

Based on Eq.~\eqref{K^mu}, we can obtain the differential equation governing the light trajectory, 
\begin{align}
    \frac{\mathrm{d}r}{\mathrm{d}\varphi} &=\pm \frac{C(r)}{b}\sqrt{\frac{1}{A(r)B(r)}-\frac{b^2}{B(r)C(r)}}
    \label{rphi}. 
\end{align}
By introducing the parameter $u\equiv 1/r$, the equation of motion of $u(\varphi)$ is obtained
\begin{align}
        \frac{\mathrm{d}u}{\mathrm{d}\varphi}&=\mp \frac{u^2 C(u)}{b}\sqrt{\frac{1}{A(u)B(u)}-\frac{b^2}{B(u)C(u)}}.
\end{align}
By tracing the light rays, Fig.\ref{light ray1} depicts the light trajectories for the SC BH, RN BH, and EMD BH. The results indicate that the presence of the magnetic charge $q$ leads to a decrease in the size of both the photon sphere and the event horizon.

\section{Spherical accretion models\label{sub3}}
Black holes themselves do not emit detectable radiation, instead, the observed emissions originate from the accreting matter surrounding them. As this material orbits the black hole, it emits light, producing observable phenomena including black hole shadows, gravitational lensing, and photon rings. This section investigates the observational signatures of black holes surrounded by optically thin spherical accretion flows. By simulating images for both static and infalling accretion scenarios, we analyze and compare this two models.

The specific intensity can be defined as in Ref.~ \cite{r54,r55,r56,r57,r58}.
\begin{align}
    I_{\mathrm{obs}}(b,\nu_o) =\int_{\mathrm{ray}} g^3 j(\nu_e)\mathrm{d}l,
    \label{Iobs}
\end{align}
where $\nu_e$ represents the frequency of photons emitted by the accreting matter, and $g$ denotes the redshift factor, given by
\begin{align}
    g = \frac{\nu_o}{\nu_e} = \frac{(-K^\mu U_\mu)|_{\mathrm{obs}}}{(-Z^\nu U_\nu)|_{\mathrm{em}}}.
    \label{g}
\end{align}
Furthermore, $\nu_o$ is the photon frequency emitted by the accreting matter, measured by a static observer at infinity. $K^\mu$ denotes the photon's four-momentum, $Z^\mu$ corresponds to the four-velocity of the accreting matter, and $U^\mu$ indicates the four-velocity of the observer. The lower index $\text{obs}$ and $\text{em}$ denote the observation and emission spacetime points of the luminous accreting matter, respectively. The lower index $\mathrm{ray}$ refers to the light trajectory, and $\mathrm{d} l$ stands for the proper distance, defined as \cite{r59}
\begin{align}
    \mathrm{d}l := -K_\mu Z^\mu \mathrm{d}\lambda,
    \label{dl}
\end{align}
$j(\nu_e)$ represents the emissivity per unit volume at frequency $\nu_e$, as measured in the local rest frame of the accreting matter. Adopting the emission model from reference \cite{r54}, we consider monochromatic emission at a rest-frame frequency $\nu$ with a radial profile that scales as $1/r^2$.
\begin{align}
    j(\nu_e) \propto \frac{\delta(\nu_e - \nu)}{r^2}.
    \label{j()}
\end{align}
Finally, by integrating over all frequencies, we obtain the integrated intensity \cite{r60}
\begin{align}
    F_{\mathrm{obs}} = \int_0^{+\infty} I_{\mathrm{obs}}(b,\nu_o) \mathrm{d}\nu_o.
    \label{Fobs}
\end{align}
\subsection{Shadows and photon rings of static spherical accretion}
From the normalization condition $g_{\mu\nu}Z^\mu Z^\nu=-1$ for the accreting matter's four-velocity, we derive $Z^\mu = \left(\dfrac{1}{\sqrt{A(r)}},0,0,0\right)$ in the static accretion model, while a stationary observer at infinity has $U^\mu = (1,0,0,0)$. Substituting Eqs.~\eqref{K^mu} and \eqref{g} then gives the redshift factor expression
\begin{align}
    g = \sqrt{A(r)}.
    \label{g1}
\end{align}
By utilizing Eqs. \eqref{K^mu}, \eqref{Iobs}, \eqref{dl}, \eqref{j()}, \eqref{Fobs}, and \eqref{g1}, the expression for the integrated intensity is obtained as
\begin{align}
    F_{\mathrm{obs}} = \int_0^{+\infty}\frac{A(r)^2}{r^2}\sqrt{\frac{B(r)C(r)}{C(r)-b^2 A(r)}}\mathrm{d}r.
\end{align}
\begin{figure}[h]                   
    \centering                      
    \includegraphics[width=1\linewidth]{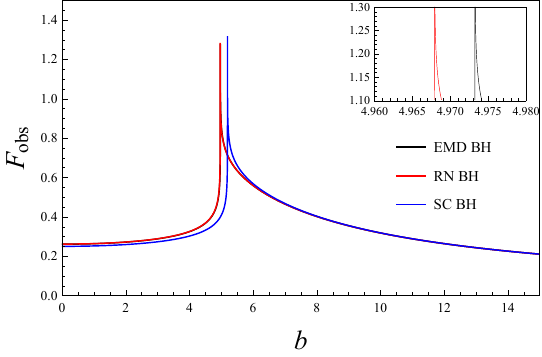}
    \caption{The distribution of the integrated intensity $F_{\text{obs}}$ with respect to the impact parameter $b$ for the three black holes under the static spherical accretion model. The blue, red, and black curves correspond to the SC BH, RN BH, and EMD BH, respectively.}                  
    \label{Fobs1}                       
\end{figure}

The Fig. \ref{Fobs1} displays the integrated intensity $F_{\text{obs}}$ as a function of impact parameter $b$ for the three black holes.  The behavior of $F_{\text{obs}}$ is qualitatively similar in all cases: it originates from a finite value and experiences a rapid increase until reaching the photon ring location $b_p$. Since the photon sphere corresponds to an unstable circular orbit for photons, light rays orbit the black hole multiple times near the photon ring, resulting in an infinite optical path length \cite{r61}. Consequently, a distant observer detects maximum intensity at the critical impact parameter, which appears as a sharp, narrow peak in Fig. \ref{Fobs1}. As $b$ increases further, the integrated intensity drops abruptly to a gentle decline, eventually converging to similar values. Notably, the photon ring manifests not as an extremum point \cite{r62} but as a sharp, narrow peak. Due to the metric similarities between EMD and RN black holes mentioned earlier, their integrated intensities are nearly identical. The inset in the upper right corner of Fig. \ref{Fobs1} provides a magnified view of the photon ring region.
\begin{figure*}[h]
    \begin{minipage}{0.32\textwidth}
        \centering
        \includegraphics[width = 0.9\textwidth]{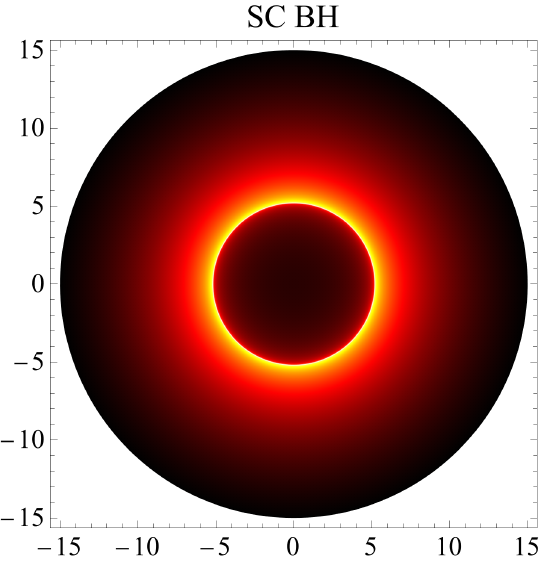}
    \end{minipage}
    \begin{minipage}{0.32\textwidth}
        \centering
        \includegraphics[width = 0.9\textwidth]{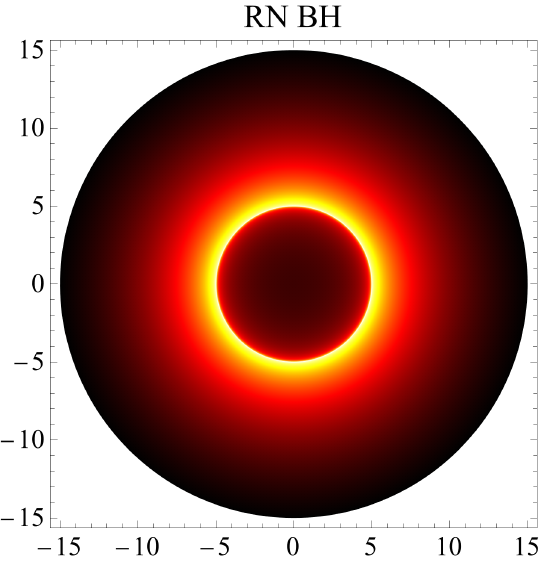}
    \end{minipage}
    \begin{minipage}{0.32\textwidth}
        \centering
        \includegraphics[width = 0.9\textwidth]{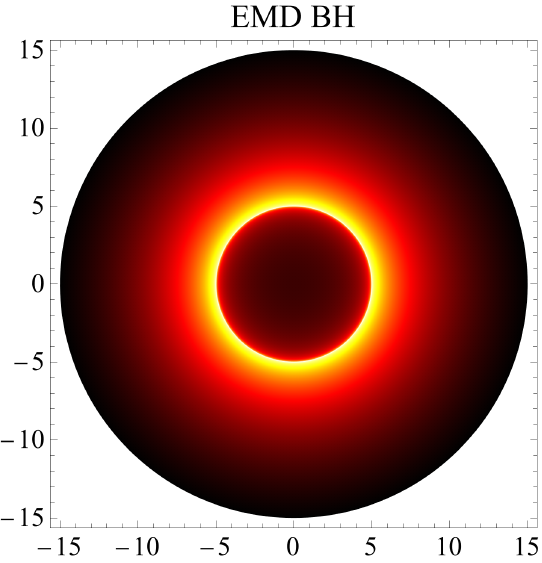}
    \end{minipage}
    \caption{{The two-dimensional images of black hole shadows and photon rings under the static spherical accretion model. From left to right: SC BH, RN BH, EMD BH.}}
    \label{Flatten Fobs1}
\end{figure*}

The Fig. \ref{Flatten Fobs1} presents the two-dimensional images in the celestial coordinate system. It can be observed that the central shadow region is surrounded by a bright photon ring. However, the intensity in the central dark region is not zero, which is due to the escape of a small number of photons from the black hole. The size of the black hole shadow and the photon ring decrease with increasing magnetic charge $q$, consistent with the conclusions in Table \ref{rhrpbp}. Furthermore, Fig. \ref{Flatten Fobs1} shows that due to the presence of hair, the brightness within the shadow region of both the EMD BH and the RN BH is higher than that of the SC BH.
\subsection{Shadows and photon rings of infalling spherical
accretion}
In the infalling accretion model, the four-velocity of a stationary observer at infinity remains $U^\mu = (1,0,0,0)$. Assuming the accreting matter starts from rest at infinity, its four-velocity becomes $Z^\mu = \left( \dfrac{1}{A(r)}, \pm \sqrt{\dfrac{1}{B(r)}\left(\dfrac{1}{A(r)}-1\right)}, 0, 0 \right)$, where the plus and minus signs correspond to light rays moving away from and toward the black hole, respectively. Consequently, in the infalling scenario, the radial component $Z^1$ of the matter's four-velocity is negative. Substitution into Equations \eqref{K^mu} and \eqref{g} then yields the expression of the red shift factor
\begin{align}
     g_{\pm} = \dfrac{A(r)}{1\pm\sqrt{1-A(r)}\sqrt{1-\dfrac{b^2 A(r)}{C(r)}}},
     \label{g2}
\end{align}
by utilizing Eqs. \eqref{K^mu}, \eqref{Iobs}, \eqref{dl}, \eqref{j()}, \eqref{Fobs}, and \eqref{g2}, the expression for the integrated intensity is obtained as
\begin{align}
     F_{obs} = \int_0^{+\infty}\frac{g_{\pm} ^4}{r^2}\left(\sqrt{\frac{B(r)C(r)}{C(r)-b^2 A(r)}} - \sqrt{\frac{B(r)(1-A(r))}{A(r)}}\right)\mathrm{d}r.
\end{align}
\begin{figure}[h]                   
    \centering                      
    \includegraphics[width=1\linewidth]{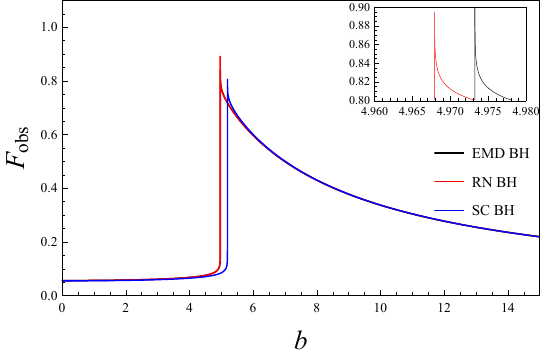}
    \caption{The distribution of the integrated intensity $F_{\text{obs}}$ with respect to the impact parameter $b$ for the three black holes under the infalling accretion model.}                  
    \label{Fobs2}                       
\end{figure}

The Fig. \ref{Fobs2} similarly demonstrates that the variation of $F_{\text{obs}}$ with different values of the parameter $q$ remains qualitatively consistent. However, compared to Fig. \ref{Fobs1}, the initial integrated intensity in Fig. \ref{Fobs2} is lower, and its rise near the photon ring location $b_p$ is more abrupt, leading to a sharper intensity contrast between the shadow's interior and exterior. A significant observation is that across different spherical accretion models, the shadow radius $b_p$ remains invariant with respect to the specific accretion scenario and is governed exclusively by the black hole's hair parameters.
\begin{figure*}[h]
    \begin{minipage}{0.32\textwidth}
        \centering
        \includegraphics[width = 0.9\textwidth]{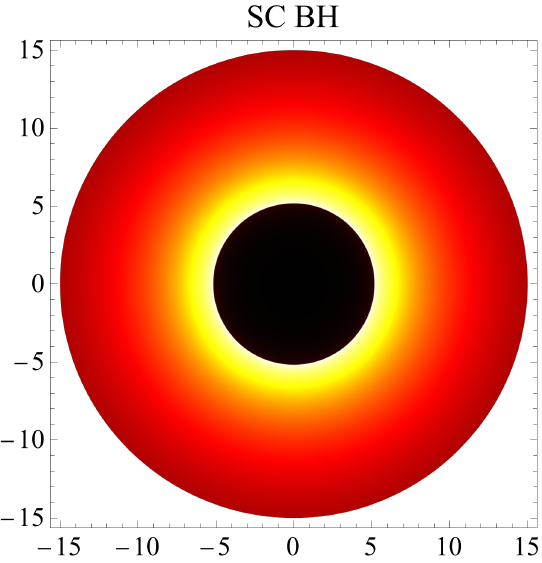}
    \end{minipage}
    \begin{minipage}{0.32\textwidth}
        \centering
        \includegraphics[width = 0.9\textwidth]{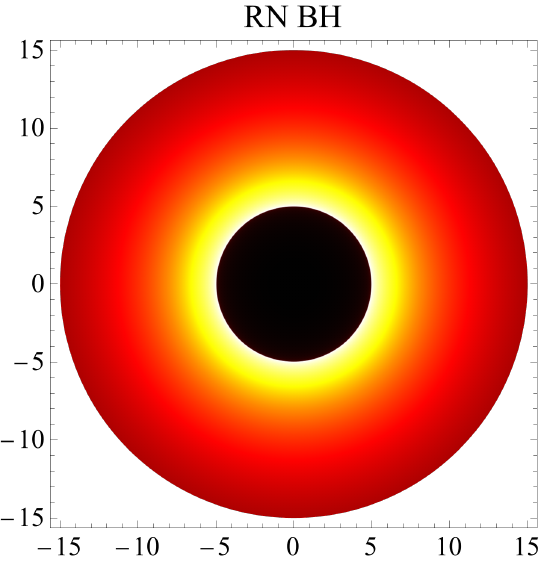}
    \end{minipage}
    \begin{minipage}{0.32\textwidth}
        \centering
        \includegraphics[width = 0.9\textwidth]{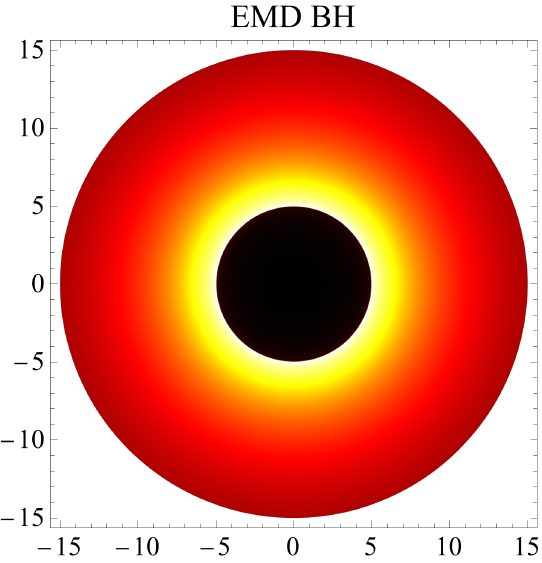}
    \end{minipage}
    \caption{{Two-dimensional images of the black hole shadow and photon ring under the infalling accretion model.}}
    \label{Flatten Fobs2}
\end{figure*}

The Fig. \ref{Flatten Fobs2} displays the two-dimensional image for the infalling accretion model, where the central dark shadow is similarly encircled by a bright photon ring. The image reveals that the shadow's size decreases with increasing magnetic charge $q$, consistent with the results in Table \ref{rhrpbp}. Moreover, the shadow region appears darker compared to Fig. \ref{Flatten Fobs1}, while the intensity outside the shadow is markedly brighter. This effect arises because the infalling accreting matter possesses a radial velocity, producing a Doppler shift that influences the observed intensity.

\section{Thin disk accretion\label{sub4}}
In this section, we consider an optically thin and geometrically thin accretion disk located in the equatorial plane, with the observer positioned at the North Pole (a face-on view of the disk). By studying light trajectories with different impact parameters, we analyze and discuss the imaging of the disk model.
\subsection{Direct emission, lensed ring emission and photon ring emission}
Photons emitted from the accretion disk orbit the black hole due to its gravitational influence and eventually reach the observer. Throughout this trajectory, the relationship between the deflection angle $\varphi$ and the number of times the photon crosses the disk is given by $\varphi = \pi/2 + (k-1)\pi$, where $k$ ($k \geq 1$) denotes the total number of intersections with the accretion disk. This formulation consequently defines the total number of photon orbits
 \begin{align}
     n = \frac{\varphi}{2\pi} = \frac{k}{2} - \frac{1}{4}.
     \label{nb equ}
 \end{align}
 
Following the methodology of reference \cite{r37}, the value obtained from Eq. \eqref{nb equ} can be used to categorize the radiation into direct emission, lensing ring emission, and photon ring emission. Direct emission ($0.25 < n < 0.75$) corresponds to light rays that cross the accretion disk only once before reaching the observer from the front side. Lensing ring emission ($0.75 < n < 1.25$) originates from photons that intersect the disk twice (passing through the thin disk) and ultimately approach the observer from the back side; during this process, the light gains additional intensity from the second passage through the accretion disk. Photon ring emission ($1.25 < n$) results from light rays that cross the disk at least three times, allowing the photons to accumulate further intensity through multiple interactions with the disk before reaching the observer \cite{r61}. Consequently, for different impact parameters (corresponding to different numbers of disk crossings), the total observed intensity comprises the contributions from these three emission types.
\begin{figure*}[h]
    \begin{minipage}{0.32\textwidth}
        \centering
        \includegraphics[width = 0.9\textwidth]{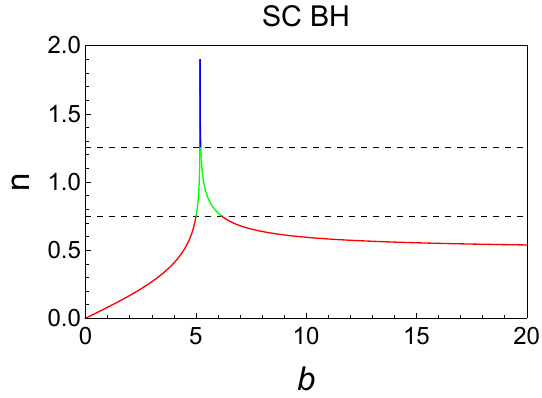}
    \end{minipage}
    \begin{minipage}{0.32\textwidth}
        \centering
        \includegraphics[width = 0.9\textwidth]{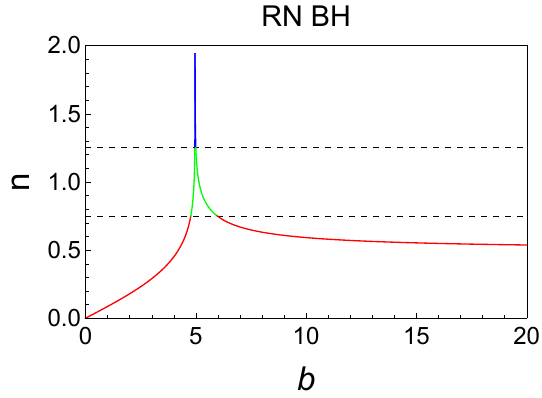}
    \end{minipage}
    \begin{minipage}{0.32\textwidth}
        \centering
        \includegraphics[width = 0.9\textwidth]{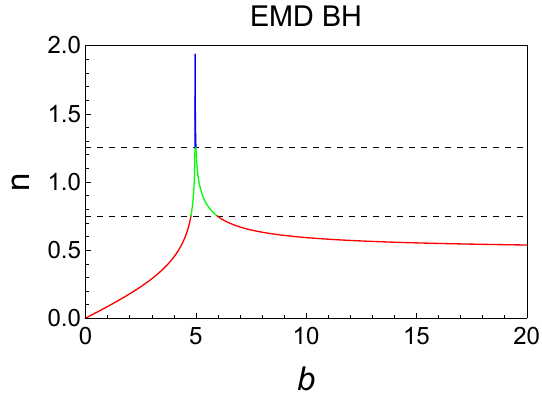}
    \end{minipage}
    \caption{{A detailed plot of the total photon orbit number $n$ versus the impact parameter $b$.}}
    \label{nbqqq}
\end{figure*}

Fig. \ref{nbqqq} displays the relationship between the total photon orbit number $n$ and the impact parameter $b$ for SC BH, RN BH, and EMD BH. The various emission types are distinguished by dashed lines in the figure, with red, green, and blue areas representing direct emission, lensing ring emission, and photon ring emission, respectively. Additionally, Table \ref{nb Table} summarizes the corresponding numerical ranges for these three emission categories.
 \begin{table*}[ht]
    \centering
    \small                        
        \caption{The specific numerical ranges corresponding to the three types of emission for the three black holes.} 
        \vspace{0.3cm}
        \begin{tabular}{c c c c}  
            \hline
            Black Hole Type \& Emission Type & Direct Emission & Lensing Ring Emission & Photon Ring Emission \\
            \hline
            SC BH   & $b$<5.01514; $b$>6.16757   & 5.01514<$b$<5.18781; 5.22794<$b$<6.16757 & 5.18781<$b$<5.22794\\
            RN BH   & $b$<4.77294; $b$>5.97448   & 4.77294<$b$<4.95793; 5.0039<$b$<5.97448 & 4.95793<$b$<5.0039 \\
            EMD BH   & $b$<4.78013; $b$>5.97631   & 4.78013<$b$<4.96345; 5.00881<$b$<5.97631 & 4.96345<$b$<5.00881 \\
            \hline
        \end{tabular}
        \label{nb Table}
\end{table*}
\begin{table*}[ht]
    \centering
    \footnotesize                  
    \caption{The thickness of the lensing ring and photon ring for the EMD black hole under different magnetic charge.}
    \vspace{0.3cm}
        \begin{tabular}{c c c c c c c c c c c c}  
            \hline
            $\Delta b$ \& Parameters & $q=0$ & $q=0.1$ & $q=0.2$ & $q=0.3$ & $q=0.4$ & $q=0.5$ & $q=0.6$ & $q=0.7$ & $q=0.8$ & $q=0.9$ & $q=1$   \\
            \hline
            $\Delta$Lensing ring   & 1.15243 & 1.15408 & 1.15907 & 1.16754 & 1.17977 & 1.19618 & 1.21738 & 1.24433 & 1.2785 & 1.32223 & 1.37955 \\
            $\Delta$Photon ring   & 0.04013 & 0.04031 & 0.04089 & 0.04188 & 0.04335 & 0.04536 & 0.04806 & 0.05164 & 0.05643 & 0.06298 & 0.07228 \\
            \hline
        \end{tabular}
        \label{detal}
\end{table*}

\begin{figure*}[h]                   
    \centering                      
    \includegraphics[width = 0.6\textwidth]{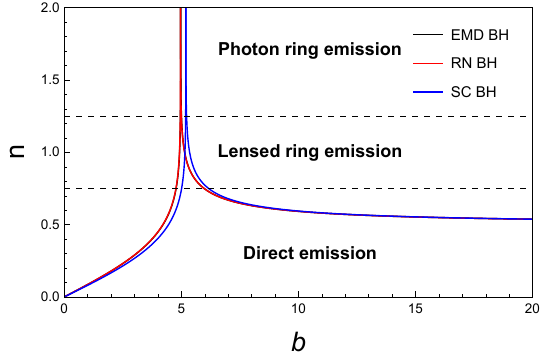}
    \caption{A comparison of the distribution of photon orbits with respect to the impact parameter for different black holes.}                  
    \label{nb Figure}                       
\end{figure*}

Table \ref{nb Table} reveals that the magnetic charge $q$ directly affects the impact parameters associated with the lensing ring and photon ring. The presence of the magnetic charge $q$ leads both rings to shift closer to the black hole, a finding consistent with the results in Table \ref{rhrpbp}. Moreover, Table \ref{detal} demonstrates how $q$ influences the widths of the lensing and photon rings. As $q$ increases, the thicknesses of both rings exhibit a positive correlation. When the impact parameter $b$ exceeds a certain threshold, the emission corresponds to direct emission regardless of the $q$ value. This indicates that both $q$ and $b$ influence the observational features of the black hole \cite{r28}. Fig. \ref{nb Figure} provides a clearer comparison of the $n(b)$ function for different $q$ values. As noted earlier, differences in the $n(b)$ function between EMD BH and RN BH only become discernible under magnification.

\begin{figure*}[h]
    \begin{minipage}{0.32\textwidth}
        \centering
        \includegraphics[width = 0.9\textwidth]{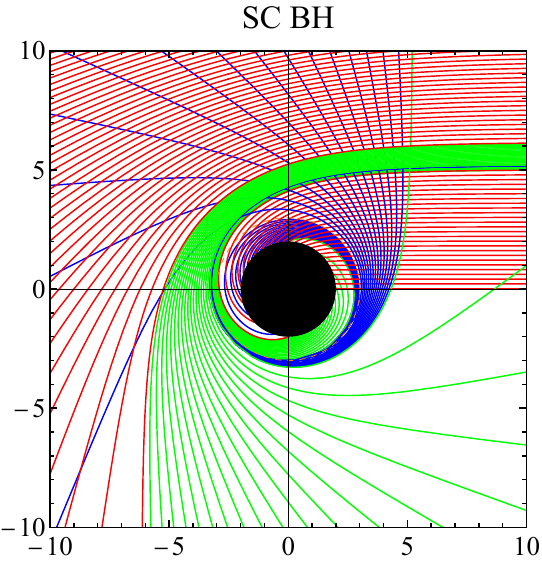}
    \end{minipage}
    \begin{minipage}{0.32\textwidth}
        \centering
        \includegraphics[width = 0.9\textwidth]{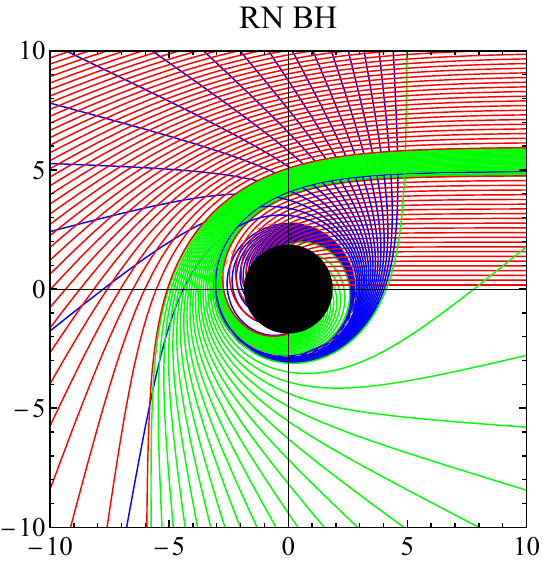}
    \end{minipage}
    \begin{minipage}{0.32\textwidth}
        \centering
        \includegraphics[width = 0.9\textwidth]{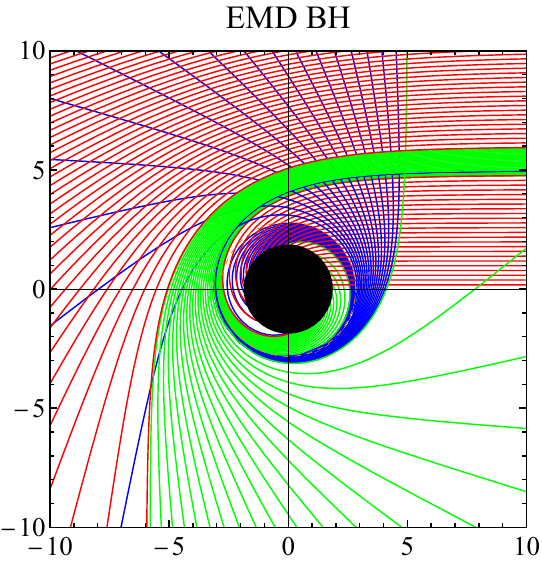}
    \end{minipage}
    \caption{{Trajectories of light rays corresponding to direct emission, lensing ring emission, and photon ring emission. The black disk represents the event horizon, the orange ring represents the photon ring. Red, green, and blue curves correspond to the photon trajectories of direct emission, lensing ring emission, and photon ring emission, respectively.}}
    \label{light ray2}
\end{figure*}

To better resolve the distribution of light trajectories near the black hole, Fig. \ref{light ray2} displays the corresponding photon paths.
\subsection{Integrated intensity and transfer function}
\begin{figure*}[h]
    \begin{minipage}{0.32\textwidth}
        \centering
        \includegraphics[width = 0.9\textwidth]{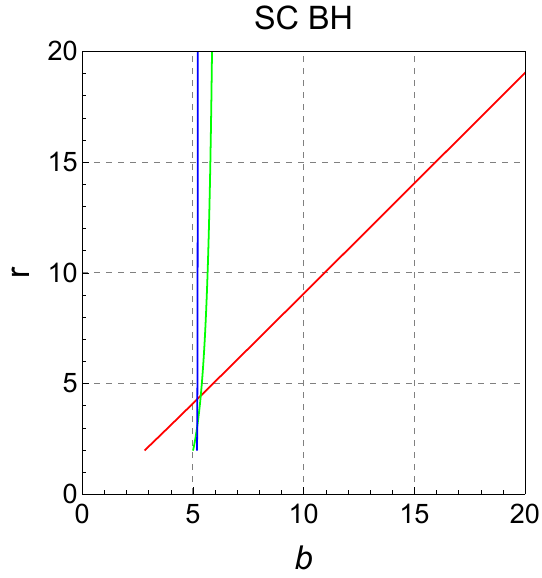}
    \end{minipage}
    \begin{minipage}{0.32\textwidth}
        \centering
        \includegraphics[width = 0.9\textwidth]{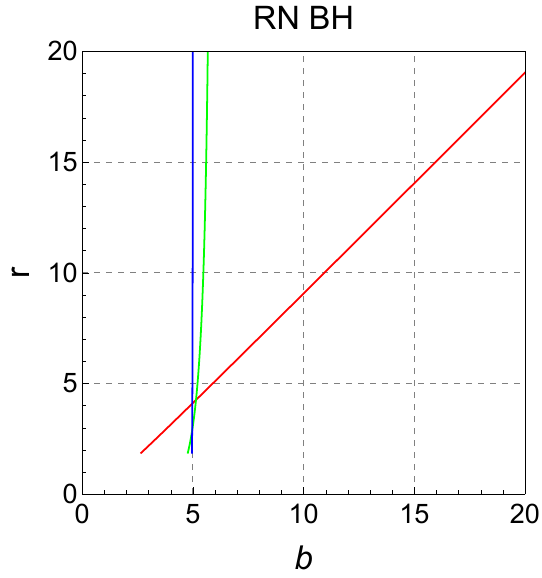}
    \end{minipage}
    \begin{minipage}{0.32\textwidth}
        \centering
        \includegraphics[width = 0.9\textwidth]{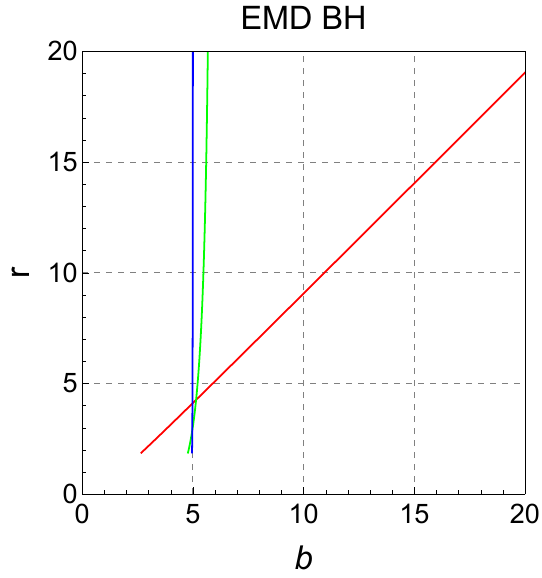}
    \end{minipage}
    \caption{{Transfer functions for different black holes. The red, green, and blue curves correspond to direct emission, lensing ring emission, and photon ring emission, respectively.}}
    \label{transfer Function}
\end{figure*}

According to Liouville's theorem, $I_e(r) /\nu_e^3$ is conserved along a light ray, therefore we have
 \begin{align}
    \frac{I_e(r)}{\nu_e^3} = \frac{I_o(r)}{\nu_o^3},\qquad
    I_o(r) = I_e(r) \left(\frac{\nu_o}{\nu_e}\right)^3 = g^3 I_e(r).
\end{align}
$I_e(r)$ represents the specific intensity of the emitted light, $\nu_e$ denotes the frequency of the emitted light, $I_o(r)$ represents the specific intensity of the received light, and $\nu_o$ denotes the frequency of the received light. To obtain the integrated intensity, it is necessary to integrate over all received frequencies
\begin{align}
    F_{\mathrm{obs}}(r) = \int_0^{+\infty} I_o(r) \mathrm{d}\nu_o = \int_0^{+\infty} g^4 I_e(r) \mathrm{d}\nu_e = A^2(r)I_{\mathrm{emit}}(r),
\end{align}
$I_{\mathrm{emit}}(r)$ represents the specific intensity of the radiation from the accretion disk, given by
\begin{align}
    I_{\mathrm{emit}}(r) = \int_0^{+\infty} I_e(r) \mathrm{d}\nu_e.
\end{align}
In the disk accretion model, photons emitted from the accreting matter eventually reach the observer. For light trajectories that cross the disk multiple times before arrival, each passage through the disk augments the observed intensity. Thus, the total received intensity corresponds to the sum of intensities from all contributing emissions.
\begin{align}
    F_{\mathrm{obs}} = \sum_k f^2(r)I_{\mathrm{emit}}|_{r = r_k(b)},
\end{align}
where $r_k(b)$ is the transfer function, which describes the radial coordinate $r$ on the disk as a function of the impact parameter $b$ and the number of intersections $k$ with the accretion disk \cite{r37}. It can be obtained from the integral result of Eq. \eqref{rphi}
\begin{align}
    r_k(b) = \dfrac{1}{u(\pi/2 + (k-1)\pi,b)}.
\end{align}
The slope of the transfer function, $\Gamma = \left.\dfrac{\mathrm{d}r}{\mathrm{d}b}\right|_k$, typically represents the magnification factor. Fig. \ref{transfer Function} shows the transfer functions for different black holes. The red curve corresponds to direct emission ($k=1$), where the slope $\left.\dfrac{\mathrm{d}r}{\mathrm{d}b}\right|_1$ remains relatively small, approximately equal to unity. This transfer function produces the direct emission image, representing the redshifted source profile. The green curve describes lensing ring emission ($k=2$), where the slope $\left.\dfrac{\mathrm{d}r}{\mathrm{d}b}\right|_2$ indicates a high magnification factor for the back side of the disk. Within this impact parameter range, alternating appearances of the front and back sides become visible. The blue curve represents photon ring emission ($k=3$), where the slope $\left.\dfrac{\mathrm{d}r}{\mathrm{d}b}\right|_3$ tends to infinity, signifying an even greater magnification of the accretion disk image.

Fig. \ref{transfer Function} further illustrates that the slope associated with lensing ring emission concentrates its resulting image within a more limited range of impact parameters. The observed intensity of this component constitutes approximately $\dfrac{1}{\Gamma}$ of the total received intensity. In contrast, the observed image of photon ring emission is confined to an even narrower interval, indicating that it should manifest as an extremely narrow ring with a negligible contribution to the overall observed intensity. Consequently, within the disk model framework, both the shadow boundary and the integrated intensity are primarily governed by direct emission, as will be evidenced in the subsequent analysis of various accretion models under this configuration.
\subsection{Different thin disk models}
We examine three different emission models under the thin disk assumption and analyze their corresponding simulated images. Each model is associated with a specific starting position for the emission, beyond which the intensity drops sharply.

For the first emission model, we consider that the accretion disk begins emitting at the innermost stable circular orbit (ISCO) of massive particles, $r_\mathrm{ISCO}$, and the total specific intensity $I_{\mathrm{emit}}$ follows a power-law decay with exponent-2 as a function of the radial coordinate
\begin{equation}
        I_{\mathrm{emit}}(r)=\begin{cases}
            \dfrac{1}{[r-(r_{\mathrm{I S C O}}-1)]^2} & r > r_{\mathrm{I S C O}}\,, \\
            0 & r \leq r_{\mathrm{I S C O}}\,.
       \end{cases}
\end{equation}
The expression for the innermost stable circular orbit $r_\mathrm{ISCO}$ of massive particles is given by \cite{r63}
\begin{align}
    r_\mathrm{I S C O} = \frac{3A(r_\mathrm{I S C O})A'(r_\mathrm{I S C O})}{2A'(r_\mathrm{I S C O})^2-A(r_\mathrm{I S C O})A''(r_\mathrm{I S C O})},
\end{align}
where $A'(r)$ and $A''(r)$ represent the first and second derivatives of $A(r)$, respectively.
\begin{figure*}[h]
    \begin{minipage}{0.32\textwidth}
        \centering
        \includegraphics[width = 0.9\textwidth]{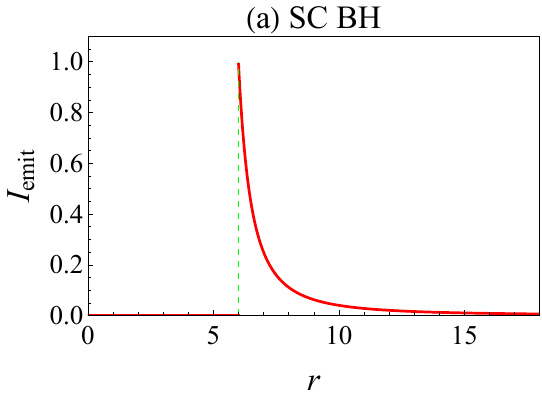}
    \end{minipage}
    \begin{minipage}{0.32\textwidth}
        \centering
        \includegraphics[width = 0.9\textwidth]{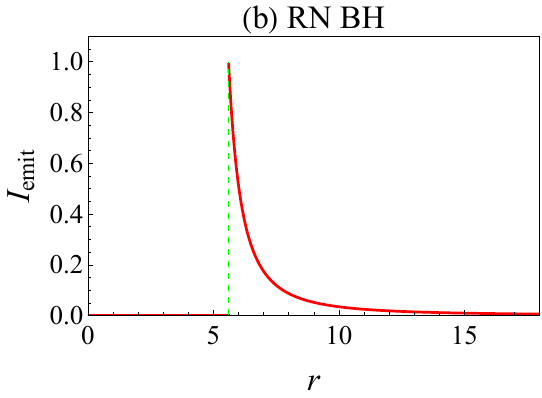}
    \end{minipage}
    \begin{minipage}{0.32\textwidth}
        \centering
        \includegraphics[width = 0.9\textwidth]{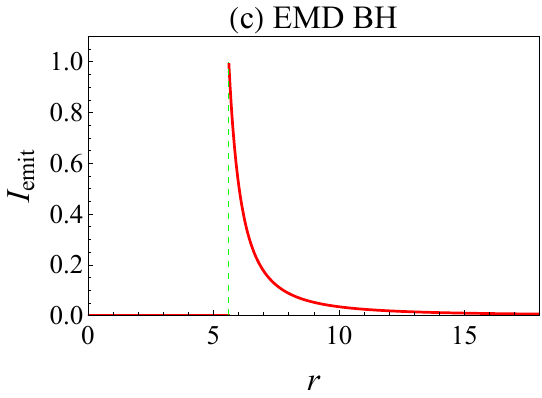}
    \end{minipage}
    \\[1em]
    \begin{minipage}{0.32\textwidth}
        \centering
        \includegraphics[width = 0.9\textwidth]{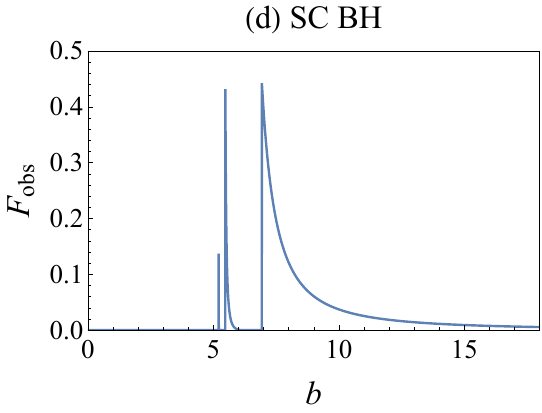}
    \end{minipage}
    \begin{minipage}{0.32\textwidth}
        \centering
        \includegraphics[width = 0.9\textwidth]{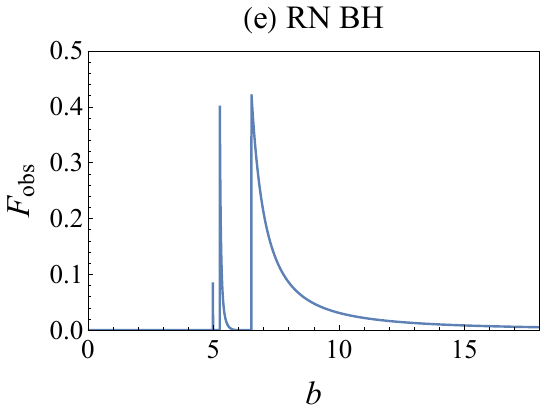}
    \end{minipage}
    \begin{minipage}{0.32\textwidth}
        \centering
        \includegraphics[width = 0.9\textwidth]{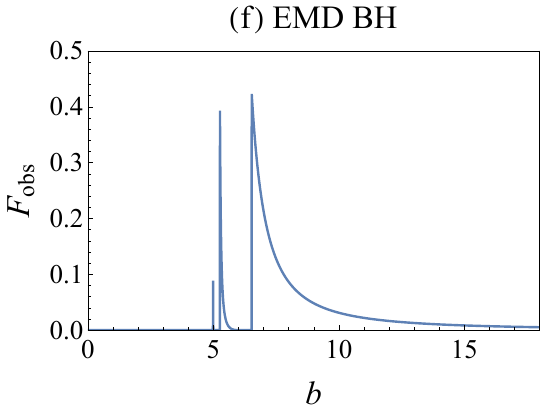}
    \end{minipage}
    \\[1em]
    \begin{minipage}{0.32\textwidth}
        \centering
        \includegraphics[width = 0.9\textwidth]{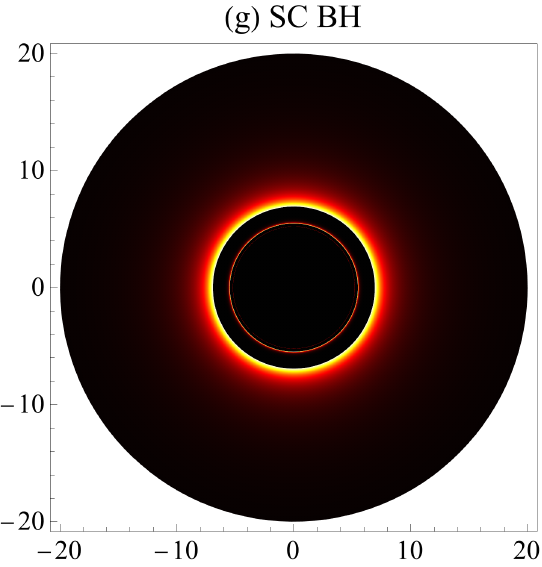}
    \end{minipage}
    \begin{minipage}{0.32\textwidth}
        \centering
        \includegraphics[width = 0.9\textwidth]{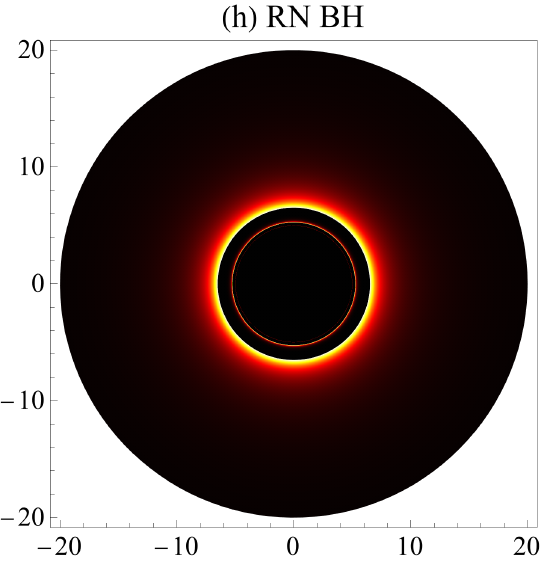}
    \end{minipage}
    \begin{minipage}{0.32\textwidth}
        \centering
        \includegraphics[width = 0.9\textwidth]{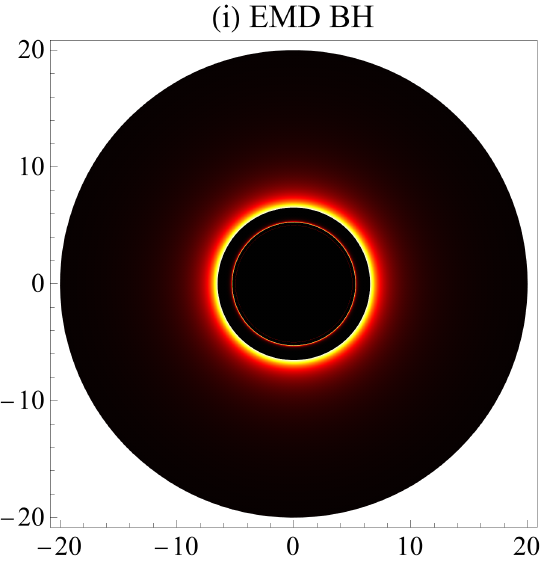}
    \end{minipage}
    \caption{The first row corresponds to the specific intensity $I_{\mathrm{emit}}$, the second row to the integrated intensity $F_{\mathrm{obs}}$, and the third row to the two-dimensional simulated images. From left to right, the columns represent the SC BH, RN BH, and EMD BH, respectively.}
    \label{r_{isco}}
\end{figure*}

Fig. \ref{r_{isco}} presents in its first row the specific intensity distribution, in the second row the observed intensity received by the observer, and in the third row the simulated images based on this observed intensity. In the first emission model, the lensing ring emission, photon ring emission, and direct emission are clearly separated. Although photons in the photon ring undergo multiple disk crossings and thus gain higher radiation intensity, their emission is confined to a narrow region. Consequently, this component appears only as a minor peak in the intensity plot and is barely discernible in the density map (third row), manifesting as an extremely fine ring visible only under magnification. While the total radiation intensity of the lensing ring exceeds that of the photon ring, its contribution amounts to only about 10\%\cite{r37} relative to direct emission. Therefore, from the observer's perspective, the lensing ring provides minimal contribution to the total brightness, the photon ring contribution is negligible, and direct emission remains the dominant source of the observed intensity.

Next, we consider the second emission model. This model begin at the photon sphere $r_p$. The specific intensity follows a cubic (third-order) power-law decay as a function of the radial coordinate
\begin{equation}
        I_{\mathrm{emit}}(r)=\begin{cases}
            \dfrac{1}{[r-(r_p-1)]^2} & r > r_p\,, \\
            0 & r \leq r_p\,.
       \end{cases}
\end{equation}

\begin{figure*}[h]
    \begin{minipage}{0.32\textwidth}
        \centering
        \includegraphics[width = 0.9\textwidth]{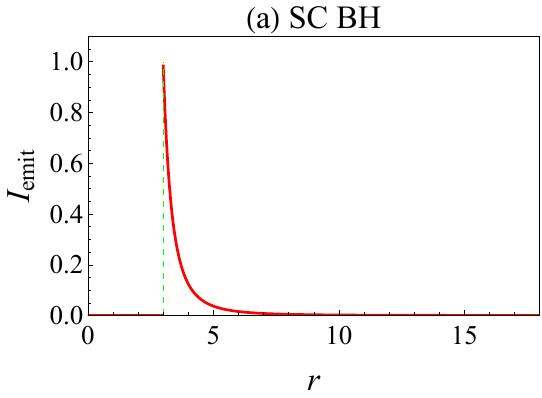}
    \end{minipage}
    \begin{minipage}{0.32\textwidth}
        \centering
        \includegraphics[width = 0.9\textwidth]{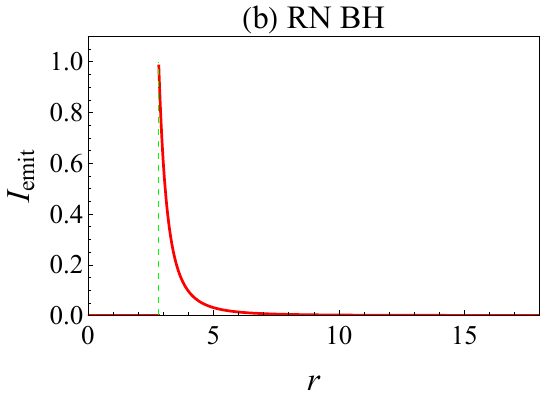}
    \end{minipage}
    \begin{minipage}{0.32\textwidth}
        \centering
        \includegraphics[width = 0.9\textwidth]{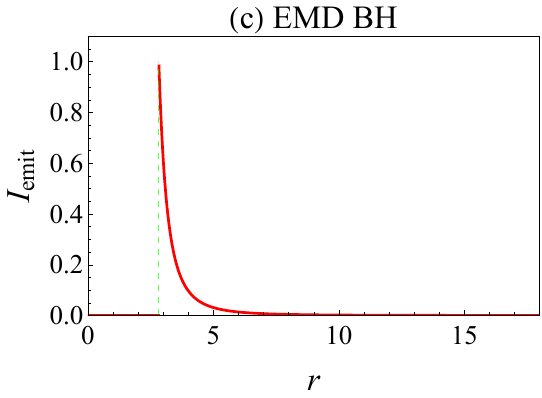}
    \end{minipage}
    \\[1em]
    \begin{minipage}{0.32\textwidth}
        \centering
        \includegraphics[width = 0.9\textwidth]{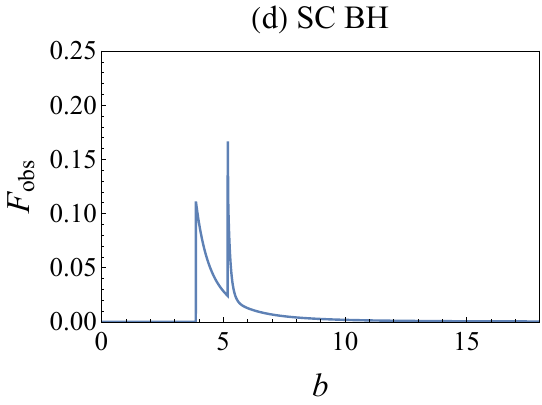}
    \end{minipage}
    \begin{minipage}{0.32\textwidth}
        \centering
        \includegraphics[width = 0.9\textwidth]{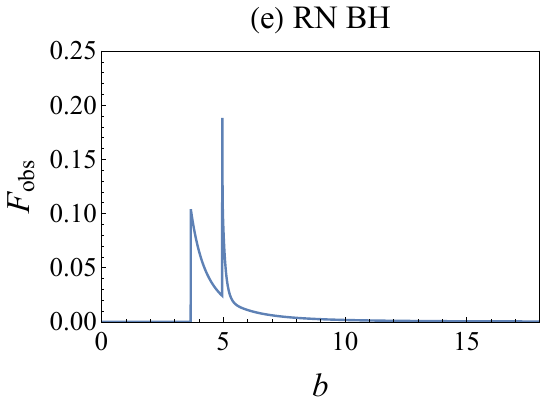}
    \end{minipage}
    \begin{minipage}{0.32\textwidth}
        \centering
        \includegraphics[width = 0.9\textwidth]{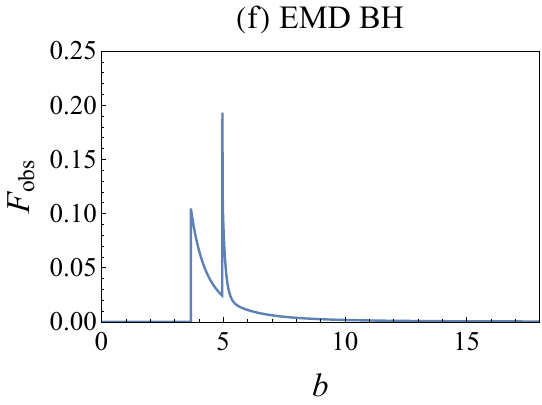}
    \end{minipage}
    \\[1em]
    \begin{minipage}{0.32\textwidth}
        \centering
        \includegraphics[width = 0.9\textwidth]{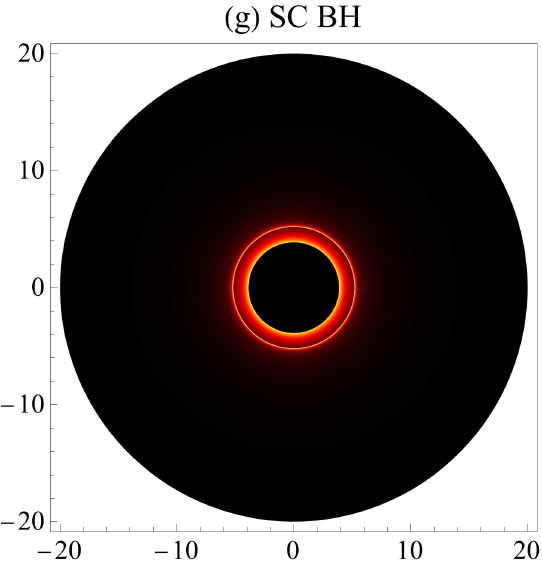}
    \end{minipage}
    \begin{minipage}{0.32\textwidth}
        \centering
        \includegraphics[width = 0.9\textwidth]{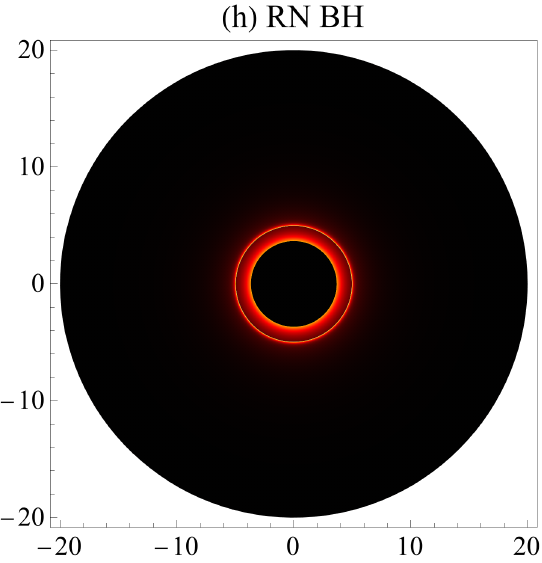}
    \end{minipage}
    \begin{minipage}{0.32\textwidth}
        \centering
        \includegraphics[width = 0.9\textwidth]{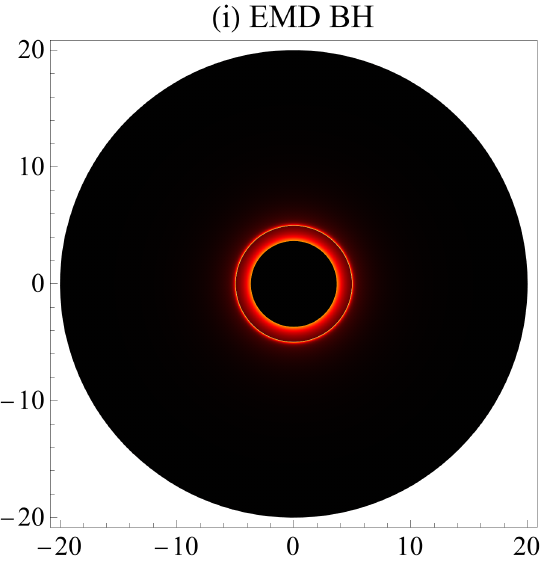}
    \end{minipage}
    \caption{Similar to Fig. \ref{r_{isco}}, the model corresponds to the second emission model.}
    \label{r_p}
\end{figure*}
Similarly, Fig. \ref{r_p} presents the observational characteristics for the second emission model. In contrast to Fig. \ref{r_{isco}}, the direct emission, lensing ring, and photon ring in Fig. \ref{r_p} overlap. This superposition makes it challenging to distinguish the individual emission components in the black hole image (third row of Fig. \ref{r_p}). Notably, the integrated intensity in Fig. \ref{r_p} is lower than in the first model, and due to the overlap of the lensing ring, photon ring, and direct emission, the observed image exhibits two bright rings. In this configuration, the shadow size is smaller, and the innermost bright ring adjacent to the shadow originates from direct emission. Ultimately, as previously indicated, the total observed intensity remains predominantly contributed by direct emission.

Finally, we consider the third emission model. This model begins at the event horizon $r_h$, and its expression is given by
\begin{align}
    I_{\mathrm{emit}}(r)=\begin{cases}
            \dfrac{\frac{\pi}{2}-\arctan(r-r_{\mathrm{ISCO}}+1)}{\frac{\pi}{2}+\arctan(r_p)} & r > r_h\,, \\
            0 & r \leq r_h\,.
       \end{cases}
\end{align}
\begin{figure*}[h]
    \begin{minipage}{0.32\textwidth}
        \centering
        \includegraphics[width = 0.9\textwidth]{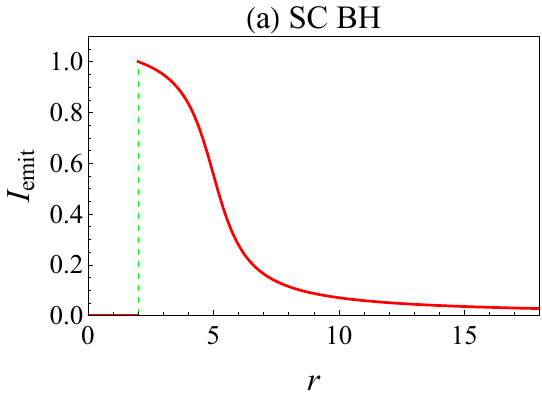}
    \end{minipage}
    \begin{minipage}{0.32\textwidth}
        \centering
        \includegraphics[width = 0.9\textwidth]{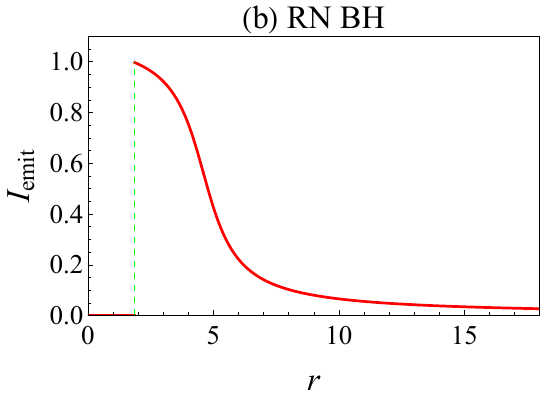}
    \end{minipage}
    \begin{minipage}{0.32\textwidth}
        \centering
        \includegraphics[width = 0.9\textwidth]{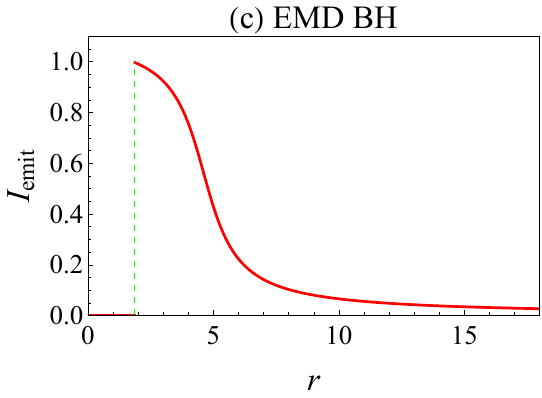}
    \end{minipage}
    \\[1em]
    \begin{minipage}{0.32\textwidth}
        \centering
        \includegraphics[width = 0.9\textwidth]{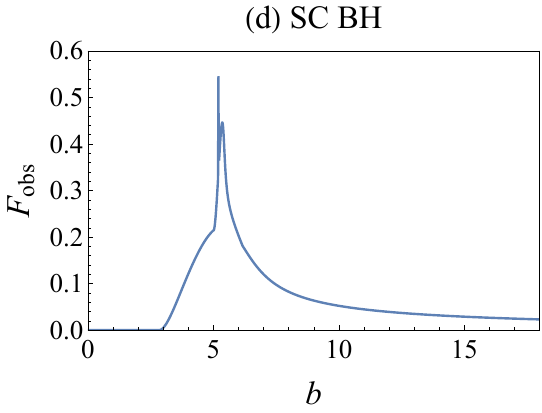}
    \end{minipage}
    \begin{minipage}{0.32\textwidth}
        \centering
        \includegraphics[width = 0.9\textwidth]{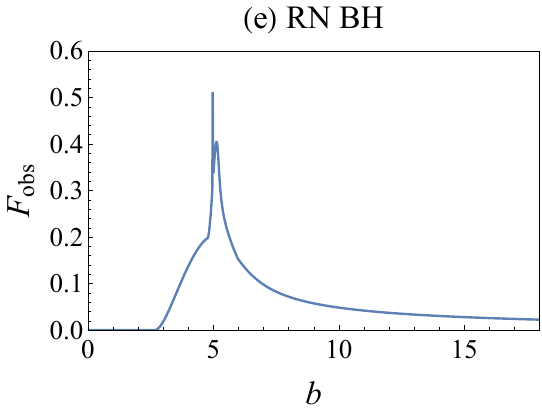}
    \end{minipage}
    \begin{minipage}{0.32\textwidth}
        \centering
        \includegraphics[width = 0.9\textwidth]{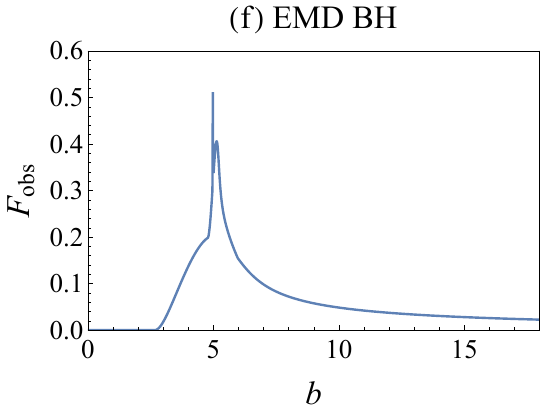}
    \end{minipage}
    \\[1em]
    \begin{minipage}{0.32\textwidth}
        \centering
        \includegraphics[width = 0.9\textwidth]{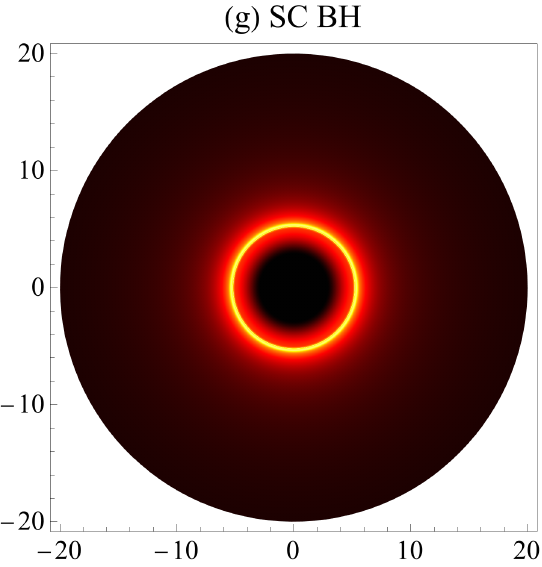}
    \end{minipage}
    \begin{minipage}{0.32\textwidth}
        \centering
        \includegraphics[width = 0.9\textwidth]{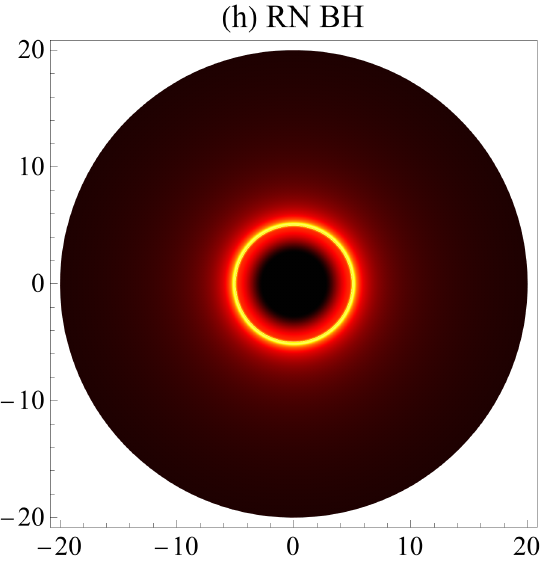}
    \end{minipage}
    \begin{minipage}{0.32\textwidth}
        \centering
        \includegraphics[width = 0.9\textwidth]{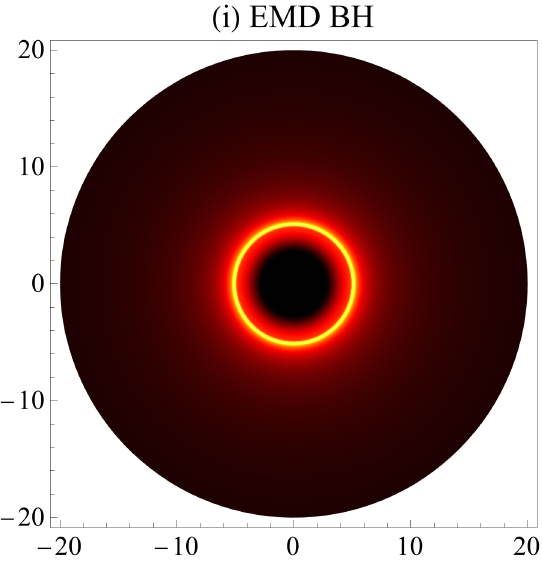}
    \end{minipage}
    \caption{Similar to Fig. \ref{r_{isco}}, the model corresponds to the third emission model.}
    \label{r_h}
\end{figure*}

Fig. \ref{r_h} displays the observational characteristics for the third emission model. In this case, the lensing ring, photon ring, and direct emission overlap and intertwine, rendering the individual emission components indistinguishable in the black hole image. This model exhibits only one bright ring, produced by the combined effect of the three emission types. Notably, as visible in the third row of Fig. \ref{r_h}, this bright ring does not define the shadow boundary; a faint luminosity persists within it, formed by direct emission at the shadow boundary. Moreover, comparison of $F_{\mathrm{obs}}$ in Figs. \ref{r_{isco}} and \ref{r_p} indicates a further reduction of the shadow region in this model.

Comparison of Figs. \ref{r_{isco}}, \ref{r_p}, and \ref{r_h} yields several important conclusions. First, regarding the relationship between $I_{\mathrm{emit}}$ and $F_{\mathrm{obs}}$, increasing distance of the initial emission location from the black hole leads to closer agreement between the overall trend of $F_{\mathrm{obs}}$ and the corresponding $I_{\mathrm{emit}}$, attributable to reduced gravitational redshift effects in distant regions. Second, across different disk emission models, the positions of both the lensing ring and photon ring remain invariant, being determined solely by the spacetime geometry. In contrast, the shadow radius varies with the emission profile; specifically, proximity of the accretion disk to the black hole results in a smaller apparent shadow radius. This represents a fundamental distinction from spherical accretion models. Furthermore, the $F_{\mathrm{obs}}$ plots demonstrate that although lensing ring and photon ring emissions possess substantially higher integrated intensities than direct emission, their narrow widths render the direct emission dominant in observations, with the photon ring contribution being negligible. Finally, the integrated intensity $F_{\mathrm{obs}}$ in the second row of Fig. \ref{r_p}, and \ref{r_h} indicate that the magnetic charge $q$ shifts the direct emission inward; compared to the SC BH, this reduces the shadow radius, consistent with the results presented in Table \ref{nb Table}.

\section{Summary and discussions\label{sub5}}
In this paper, we systematically examine light trajectories, intensity distributions, and optical imaging of EMD black hole. We first analyze how the magnetic charge $q$ modifies the spacetime geometry, subsequently affecting key features including the event horizon, photon sphere, and photon ring. Using observational constraints from the EHT , we determine the allowable range for $q$: the upper limit is approximately 0.82 at the $1\sigma$ confidence level and about 0.99 at $2\sigma$. Finally, with $q$ set to 0.5, we compare the optical characteristics of the EMD BH with those of the SC BH and RN BH (with $Q=0.5$) across various emission models.

Under the spherical accretion framework, we examine two distinct scenarios: static accretion and infalling accretion. In the static case, the central shadow region does not appear completely dark. For the infalling scenario, the Doppler effect causes the observed intensity within the shadow to be markedly lower than in the static case, producing a sharper contrast between the shadow's interior and exterior. Across both spherical accretion models, the maximum observed intensity consistently occurs near the photon sphere. Furthermore, the sizes of both the photon ring and the shadow remain invariant under model variations, being exclusively governed by the underlying spacetime geometry.

For the geometrically thin, optically thin disk accretion model, we perform a detailed analysis of how different parameters affect various emission orders. The results show that EMD BH display narrow photon and lensing rings, whose thicknesses increase with the parameter $q$. Light rays in the photon ring region can cross the accretion disk three or more times, thereby acquiring additional intensity. However, due to the extremely narrow width of the photon ring, its image is highly magnified, and its contribution to the total observed intensity is generally negligible. Additionally, images in the lensing ring region are also magnified and, from the observer’s viewpoint, exhibit an interwoven pattern of the front and back sides of the disk. Across all three emission models, direct emission dominates the total observed intensity. Notably, within the disk accretion framework, the location of the photon ring depends only on the spacetime geometry (parameter $q$) and is independent of the specific emission model. In contrast to the spherical accretion case, however, the shadow radius in the disk accretion model varies with the emission model.Throughout the article, it is demonstrated that the optical characteristics of EMD BH and RN BH are highly coincident; their distinct features can only be identified through magnification. As of now, technological limitations prevent us from distinguishing these two types of black holes based on the images we have captured.

The limits on the magnetic charge parameter $q$ derived from black hole shadow observations serve as an important observational test of EMD gravity. While current imaging facilities such as the Event Horizon Telescope can only place preliminary constraints on $q$, future advances in multi‑messenger astronomy, especially gravitational wave observations of black hole mergers, are expected to yield tighter and more robust bounds. Furthermore, a combined analysis of shadow size, photon ring structure, and gravitational wave signals will help to break degeneracies among different black hole parameters and offer deeper insight into the role of the magnetic charge $q$ in modified gravity.

\vspace{\baselineskip}
\setlength{\parindent}{0pt}\textbf{\textbf{Acknowledgments}}
This work is supported by National Key
R\&D Program of China (2024YFA1611700), Guangxi Key Research and Development Programme(Guike FN2504240040), the National Natural Science Foundation of China (Grants Nos. 12105039, 12133003, and 12494574) and the Guangxi Talent Program (``Highland of Innovation Talent'').

\appendix
\setlength{\parindent}{1.5em}

\end{document}